\documentstyle[twoside,fleqn,espcrc2,epsfig]{article}

\title{Random-matrix universality in the small-eigenvalue spectrum
  of the lattice Dirac operator}

\author{M.E.\ Berbenni-Bitsch\address{Fachbereich Physik, Universit\"at
    Kaiserslautern, D-67663 Kaiserslautern, Germany},
  A.D.\ Jackson\address{Niels-Bohr-Institute, Blegdamsvej 17, DK-2100
    Copenhagen \O, Denmark}, S.\ Meyer$^{\rm a}$,
  A.\ Sch\"afer\address{Institut f\"ur Theoretische Physik,
    Universit\"at Regensburg, D-93040 Regensburg, Germany},
  J.J.M.\ Verbaarschot\address{Department of Physics, State University
    of New York, Stony Brook, NY 11794, USA}, and
  T.\ Wettig\address{Institut f\"ur Theoretische Physik, Technische
    Universit\"at M\"unchen, D-85747 Garching,
    Germany}\thanks{Talk presented by T.\ Wettig}}
       
\begin{document}

\begin{abstract}
  We analyze complete spectra of the lattice Dirac operator in
  SU(2) gauge theory and demonstrate that the distribution of
  low-lying eigenvalues is described by random matrix theory.  We
  present possible practical applications of this random-matrix
  universality.  In particular, reliable extrapolations of lattice 
  gauge data to the thermodynamic limit are discussed.
\end{abstract}

\maketitle

\section{Introduction}
\label{sec1}

Based on an analysis of sum rules derived by Leutwyler and Smilga
\cite{Leut92} for inverse powers of the eigenvalues of the QCD
Dirac operator in a finite volume, Shuryak and Verbaarschot
\cite{Shur93} conjectured that the so-called microscopic spectral
density of the Dirac operator, $\rho_s(z)$, is a universal
function which can be computed in random matrix theory (RMT).  This
quantity is defined by
\begin{equation}
  \label{eq1.1}
  \rho_s(z)=\lim_{V\rightarrow\infty}\frac{1}{V\Sigma}
  \rho(\frac{z}{V\Sigma}) \:,
\end{equation}
where $\rho(\lambda)=\langle\sum_n\delta(\lambda-\lambda_n)\rangle_A$
is the spectral density of the Dirac operator averaged over all gauge
field configurations $A$, $V$ is the space-time volume, and $\Sigma$
is the absolute value of the chiral condensate.  The chiral condensate
can be determined using the Banks-Casher relation \cite{Bank80}
\begin{equation}
  \label{eq1.2}
  \langle\bar\psi\psi\rangle=\lim_{m\rightarrow 0}
  \lim_{V\rightarrow\infty} \frac{\pi\rho(0}{V} \:.   
\end{equation}
The definition of Eq.~(\ref{eq1.1}) amounts to a
magnification of the region of low-lying eigenvalues by a factor of
the volume.  Since the spacing of small eigenvalues is $\sim
1/(V\Sigma)$, this definition leads to the resolution of individual 
eigenvalues.

A number of studies have presented evidence supporting the
universality conjecture for $\rho_s$.  They are summarized in
Ref.~\cite{Berb97}.  Here, we present and analyze the results of a
high-statistics lattice study in SU(2) gauge theory with staggered
fermions which confirms this conjecture in a particularly direct way.
Most of the details omitted here, in particular regarding the lattice
calculations, can be found in Ref.~\cite{Berb97}, in the talk by S.\ 
Meyer, and in the poster by M.E.\ Berbenni-Bitsch, respectively (these
proceedings).

\section{Microscopic universality}
\label{sec2}

Depending on the number of colors and on the representation of the
fermions, one must distinguish three different universality classes
which were classified in Ref.~\cite{Verb94}.  These correspond to the
three chiral ensembles of RMT, the chiral Gaussian orthogonal (chGOE),
unitary (chGUE), and symplectic (chGSE) ensemble.  Analytic random-matrix 
results have been obtained for all three ensembles for a variety of 
quantities including the microscopic spectral density, the
distribution of the smallest eigenvalue, and microscopic spectral
correlators.  We have worked with staggered fermions in SU(2) for
which the appropriate RMT ensemble is the chGSE.  RMT results for
the chGSE can be obtained by a slight modification of the analytic 
results obtained by Forrester \cite{Forr93} and by Nagao and Forrester 
\cite{Naga95} for generalized Laguerre ensembles.  
\begin{figure*}[t]
\centerline{\psfig{figure=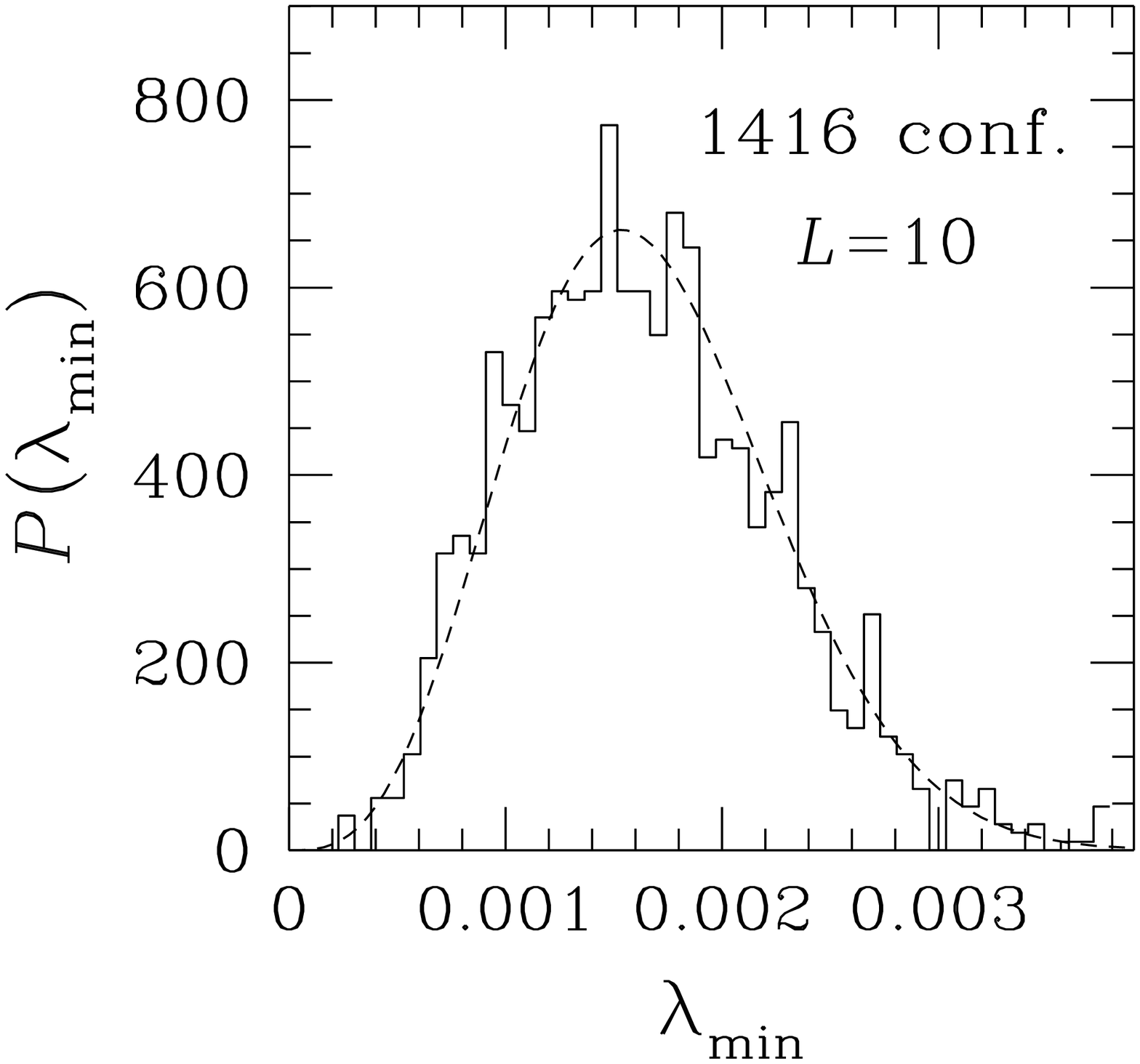,width=50mm}\hspace*{5mm}
  \psfig{figure=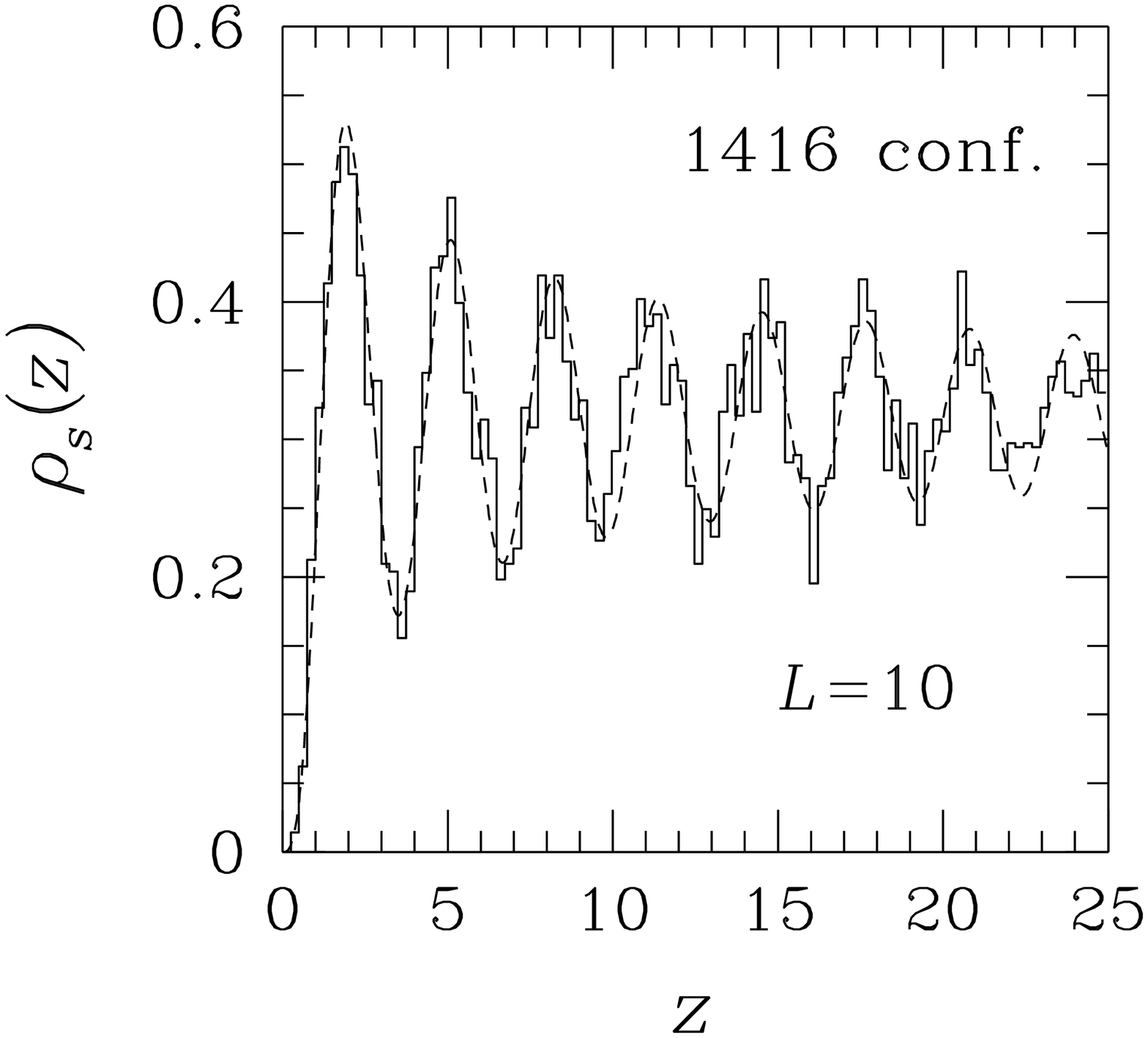,width=50mm}\hspace*{5mm}
  \psfig{figure=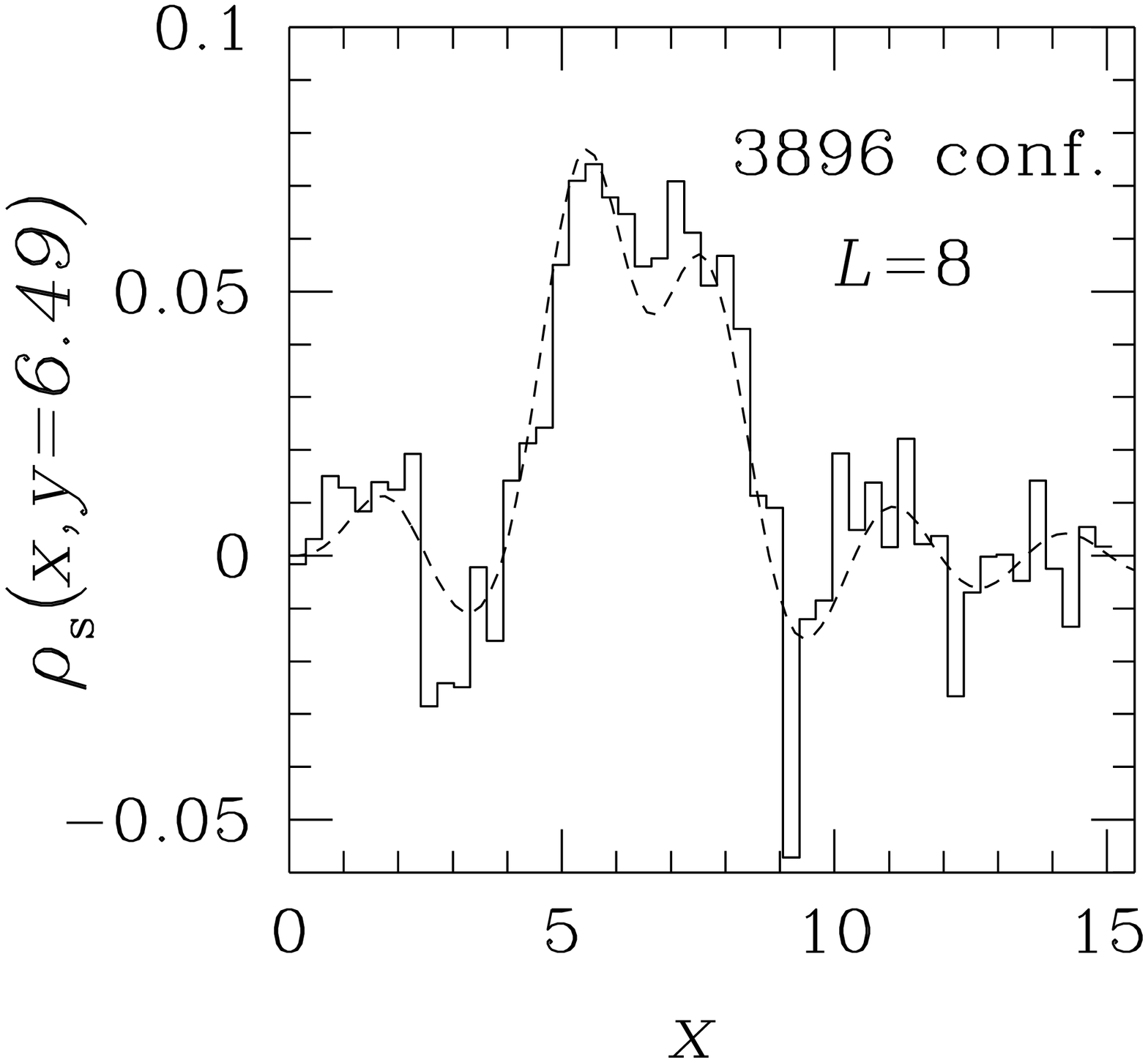,width=50mm}}
\vspace*{-10mm}
\caption{The distribution of the smallest eigenvalue $P(\lambda_{\rm
    min})$ and microscopic spectral density $\rho_s(z)$ (for a 
  10$^4$ lattice) and the microscopic spectral two-point function
  $\rho_s(x,y)$ (for an 8$^4$ lattice.)  The histograms represent
  lattice data; the dashed lines are analytical predictions from
  random matrix theory.}
\label{fig1}
\end{figure*}
In particular, we have
\begin{eqnarray}
  \rho_s(z)&\!\!\!=\!\!\!&z[J_0^2(2z)+J_1^2(2z)]\nonumber\\
  & & -\frac{1}{2}J_0(2z)\int_0^{2z} dt J_0(t)
  \label{eq2.1}
\end{eqnarray}
for the microscopic spectral density and
\begin{eqnarray}
  &&\hspace*{-20pt}P(\lambda_{\rm min})=\sqrt{\frac{\pi}{2}}c(c\lambda_{\rm
    min})^{3/2} I_{3/2}(c\lambda_{\rm min})
  e^{-\frac{1}{2}(c\lambda_{\rm min})^2} \nonumber \\ 
  \label{eq2.2}
\end{eqnarray}
for the distribution of the smallest eigenvalue.  Here, $J$ (and $I$)
are (modified) Bessel functions, and
$c=V\langle\bar\psi\psi\rangle$.  Note that Eq.~(\ref{eq2.1}) is
simpler than the equivalent expression quoted in
Ref.~\cite{Berb97}.  The RMT result for the connected microscopic
spectral two-point correlator is
\begin{equation}
  \label{eq2.3}
  \rho_s(x,y) = f\partial_x\partial_y f -\partial_x f\partial_y f \:,
\end{equation}
where
\begin{displaymath}
  f(x,y)=\frac{y}{2}\int_0^{2x}\!\!dt\,C(t,2y)
  -\frac{x}{2}\int_0^{2y}\!\!dt\,C(t,2x)
\end{displaymath}
with
\begin{displaymath}
  C(x,y) = \frac{xJ_1(x)J_0(y)-yJ_0(x)J_1(y)}{x^2-y^2} \:.
\end{displaymath}
We have performed lattice simulations using $\beta=2.0$ for four
different lattice sizes $L^4$ with $L=4,6,8,10$.  Lattice data for all
three quantities are compared with the RMT predictions of
Eqs.~(\ref{eq2.1})--(\ref{eq2.3}) in Fig.~\ref{fig1} for selected
lattice sizes.  The number of configurations is indicated in the figure.  
The agreement is remarkably good even for such modest lattice sizes.  Note 
that there are no free parameters.  The only parameter which enters the RMT 
predictions is given by the volume and the chiral condensate, which has 
been fixed by the lattice data using the Banks-Casher relation.

Since the RMT results are obtained in the thermodynamic limit, the
quality of the agreement between lattice data and RMT increases with
increasing physical volume.  Hence, for larger values of $\beta$, one
requires larger lattices to attain the same level of agreement.  We
have confirmed this expectation using $\beta=2.2$ on $6^4$ and $8^4$
lattices and $\beta=2.5$ on a $16^4$ lattice, respectively.

\section{Practical applications}
\label{sec3}

\begin{figure}[h]
\centerline{\psfig{figure=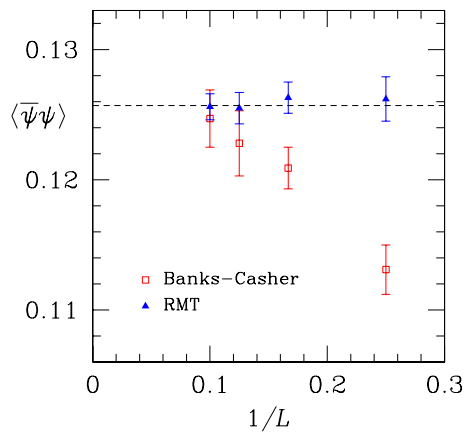,width=60mm}}
\vspace*{-5mm}
\caption{Extrapolation of $\langle\bar\psi\psi\rangle$ to the
  thermodynamic limit (indicated by the dashed line), via the
  Banks-Casher relation (boxes) and by fitting the lattice data for
  $\rho_s(z)$ to Eq.~(\protect\ref{eq2.1}) (triangles).}
\label{fig2}
\end{figure}

Having convinced ourselves and, hopefully, the reader that the
distribution of low-lying eigenvalues in the limit of large $V$ 
is given exactly by RMT, we now ask how one can make practical use 
of this knowledge.  One immediate application is the following.  We 
have pointed out earlier that the RMT predictions are parameter-free; 
the only parameter appearing in the expressions (essentially the chiral 
condensate) is fixed by the lattice data.  Given the agreement between 
lattice data and RMT, one can turn the argument around and determine the 
chiral condensate by adjusting the energy scale in order to fit the 
lattice data for, say, the microscopic spectral density to the RMT 
prediction of Eq.~(\ref{eq2.1}).  This procedure yields an independent 
estimate of the chiral condensate which should be compared to the value 
obtained from the Banks-Casher relation.  This comparison is made in 
Table~\ref{table1} where we also include the $\chi^2$ per degree of freedom 
for the fit to Eq.~(\ref{eq2.1}).
\begin{table}[t]
\setlength{\tabcolsep}{1.5pc}
\newlength{\digitwidth} \settowidth{\digitwidth}{\rm 0}
\catcode`?=\active \def?{\kern\digitwidth}
\caption{The chiral condensate as a function of the lattice size,
  determined by the Banks-Casher relation (BC) and by a fit to the
  random-matrix prediction (RMT).  See also Fig.~\protect\ref{fig2}.}
\label{table1}
\begin{tabular*}{\columnwidth}{@{\hspace*{7pt}}c@{\hspace*{20pt}}
    c@{\hspace*{20pt}}c@{\hspace*{14pt}}c@{\hspace*{5pt}}}
\hline \\[-3mm]
$L$ & $\langle\bar\psi\psi\rangle$ (BC) & 
$\langle\bar\psi\psi\rangle$ (RMT) & $\chi^2$/dof \\[1mm] \hline \\[-3mm]
$4$ & 0.1131(19) & 0.1262(17) & 6.75 \\
$6$ & 0.1209(16) & 0.1263(12) & 1.87 \\
$8$ & 0.1228(25) & 0.1255(12) & 1.78 \\
$10$ & 0.1247(22) & 0.1256(10) & 1.15 \\[1mm] \hline
\end{tabular*}
\end{table}
The same numbers are displayed in Fig.~\ref{fig2}.
The results are striking.  For all lattice sizes considered, the values
of $\langle\bar\psi\psi\rangle$ obtained from the fit to RMT agree
with each other and with the thermodynamic limit within error bars.
The values obtained from the Banks-Casher relation approach
the thermodynamic limit more slowly. The surprising result of this
analysis is that with a better accounting of finite size effects with
random matrix theory, the chiral condensate is remarkably volume 
independent. This approach appears 
to offer an interesting technical advance by making it possible to 
extract certain information about the thermodynamic limit from small 
lattices.  One can also extract scalar susceptibilities from a similar 
analysis of spectral two-point functions; work in this direction is in 
progress.  Precise numbers for the condensate 
and the susceptibilities as a function of temperature are needed to 
determine critical exponents close to the chiral phase transition.  We 
are currently testing how the proposed approach works at finite temperatures, 
in particular close to the critical temperature.

\section{Summary}
\label{sec4}

We believe to have demonstrated unambiguously that the distribution of
low-lying eigenvalues of the Dirac operator is described by RMT.  Some
practical applications of this fact have been discussed in
Sec.~\ref{sec3}.  We are confident that more can be found.

\end{document}